\documentclass[sort&compress,preprint,times,12pt]{elsarticle}%elsarticle
\usepackage{float} 
\usepackage{subfigure}  % use for side-by-side figures
\usepackage{amsmath,graphicx,url}
\newcommand{\bi}{\mathbf{i}}
\newcommand{\bj}{\mathbf{j}}
\newcommand{\bk}{\mathbf{k}}

\begin{document}
\title{A hierarchy of self-consistent stochastic boundary conditions
for Ising lattice simulations}

\author[pap]{Yidan Wang}
\author[pap,stony]{You Quan Chong}
\author[pap]{Siew Ann Cheong\corref{cor1}}
\cortext[cor1]{Corresponding author}
\ead{cheongsa@ntu.edu.sg}

\address[pap]{Division of Physics and Applied Physics,
School of Physical and Mathematical Sciences,
Nanyang Technological University,
21 Nanyang Link, Singapore 637371,
Republic of Singapore}

\address[stony]{Physics and Astronomy Department, Stony Brook University, Stony Brook, NY 11794-3800, United States of America}

\begin{abstract}
We describe a hierarchy of stochastic boundary conditions (SBCs) that can be
used to systematically eliminate finite size effects in Monte Carlo simulations
of Ising lattices.  For an Ising model on a $100 \times 100$ square lattice, we
measured the specific heat, the magnetic susceptibility, and the spin-spin
correlation using SBCs of the two lowest orders, to show that they compare
favourably against periodic boundary conditions (PBC) simulations and analytical
results.  To demonstrate how versatile the SBCs are, we then simulated an Ising
lattice with a magnetized boundary, and another with an open boundary, measuring
the magnetization, magnetic susceptibility, and longitudinal and transverse
spin-spin correlations as a function of distance from the boundary.
\end{abstract}


\maketitle

\section{Introduction}

We have learnt a great deal about the statistical physics of critical phenomena
from Monte Carlo simulations of the Ising model.  Starting with early works
\cite{Fosdick1963MethodsinComputationalPhysics, Harris1964PhysRevLett13p158,
Binder1968PhysLettA27p247} (see Ref. \cite{Binder1985JCompPhys59p1} for a
review), Monte Carlo simulations have helped us verified exact solutions for the
two-dimensional Ising model \cite{Onsager1944PhysRev65p117,
Yang1952PhysRev85p808} (see Ref. \cite{Binder1997RepProgPhys60p487} for a
review) as well as renormalization-group calculations for higher-dimensional
Ising models \cite{Fisher1972PhysRevLett29p917, Pelissetto2002PhysRep368p549}
(see Ref. \cite{Binder2001PhysRep344p179} for a review).  Associated with the
growing interest in Monte Carlo methods within the statistical and computational
physics communities, there were also notable developments of important
algorithms to tackle the problem of \emph{critical slowing down}, where the
dynamical time scale diverges as we approach the critical point
\cite{Stoll1973PhysRevB8p3266, MullerKrumbhaar1973JStatPhys8p1,
Hohenberg1977RevModPhys49p435}.  As simple Monte Carlo algorithms become highly
inefficient, many algorithmic acceleration methods based on resampling or
cluster moves have been proposed \cite{Bortz1975JCompPhys17p10,
Williams1984JStatPhys37p283, Swendsen1987PhysRevLett58p86,
Creutz1987PhysRevD36p515, Ferrenberg1988PhysRevLett61p2635,
Wolff1989PhysLettB228p379, Wolff1989PhysRevLett62p361, Swendsen1992TAP71p75,
Evertz1993PhysRevLett70p875, Lee1993PhysRevLett71p211, Li1995PhysRevLett74p3396,
Machta1995PhysRevLett75p2792, Dall2001CompPhysComm141p260,
Frenkel2004PNAS101p17571}.  A good review of the basic principles behind
acceleration algorithms can be found in Ref.~\cite{Sokal1997NATOASIB361p131}.
Recently, there have also been developments in parallelization acceleration
\cite{Lubachevsky1988JCompPhys75p103} and hardware acceleration
\cite{Talapov1996JPhysAMathGen29p5727, Preis2009JCompPhys228p4468,
Hawick2009IntJParProg}.  These represent improvements beyond the previous state
of the art in Monte Carlo simulation of Ising models
\cite{Ferrenberg1991PhysRevB44p5081, Blote1995JPhysAMathGen28p6289,
Caselle1998NuclPhysB63p613}.

The origin of this diverging dynamical time scale scale at the critical
temperature is the diverging correlation length scale
\cite{VanHove1954PhysRev93p268,VanHove1954PhysRev95p249}.  Existing acceleration
algorithms addresses only the problem of the diverging time scale.  To address
the problem of the diverging length scale, we either simulate a system whose
size is larger than the diverging length scale, or perform finite size scaling
analysis to eliminate finite size effects \cite{Binder1972Physica62p508,
Binder1981PhysRevLett47p693, Binder1989JStatPhys55p87,
Binder1992LecNotesPhys409p59, Privman1990FiniteSizeScaling,
Promberger1995ZeitsPhysB97p341, Caracciolo1995PhysRevLett74p2969}.  In nearly
all cases, we follow suggestions in texts on Monte Carlo methods
\cite{Newman1999MonteCarloMethodsinStatisticalPhysics,
Binder2002MonteCarloSimulationsinStatisticalPhysics,
Landau2009AGuidetoMonteCarloSimulationsinStatisticalPhysics}, to impose periodic
boundary conditions (PBCs).  In other studies, artificial correlations
introduced by PBCs have been observed to strongly influence the results of
computational studies \cite{Gabay1985JPhysFrance46p5, Gabay1986PhysRevB33p6281,
Kashuba1996PhysRevLett77p2554}.

In this paper, we propose a hierarchy of self-consistent stochastic boundary
conditions (SBCs) for the Monte Carlo simulation of Ising lattices.  The idea of
designing boundary conditions that allows us to minimize, or completely
eliminate finite size effects is not new.  For exact diagonalization studies or
quantum Monte Carlo simulations, \emph{twist boundary conditions} have been
introduced \cite{Spronken1981PhysRevB24p5356, Zotos1990PhysRevB42p8445,
Kohn1964PhysRev133pA171}.  In the Monte Carlo literature, Binder and M\"uller-Krumbhaar first used \emph{self-consistent boundary conditions} in their study of a simple classical Heisenberg ferromagnet \cite{MullerKrumbhaarZeitschrift254p269}.  In this boundary condition, a uniform magnetic field is applied to the boundary, and iteratively equilibriated with the average magnetization of the interior spins.  Hasenbusch then introduced \emph{fluctuating boundary conditions}, for his Monte Carlo study of the three-dimensional Ising model \cite{Hasenbusch1993PhysicaA197p423}, by averaging over periodic and antiperiodic boundary conditions.  These fluctuating boundary conditions were generalised \cite{Saslow1992PhysRevLett68p3627, Benakli1998EuroPhysJB1p197}, by introducing phase shifts in boundary spins of quantum lattices.  These boundary phase shifts are adjusted iteratively, until measurements at the boundaries are consistent with measurements within the bulk.  In addition, the name \emph{self-consistent boundary conditions} has also been used by Olsson \cite{Olsson1994PhysRevLett73p3339, Olsson1995PhysRevB52p4511} to describe his method of interpolating between periodic and fluctuating boundary conditions to achieve self-consistency between the boundary and the bulk in simulations of the XY model.  Our contributions in this paper is the development of boundary conditions that are stochastic and self-consistent, and with no freedom to perform optimization.  Finite size effects can be systematically eliminated by going to higher and higher order in the hierarchy of stochastic boundary conditions.  More importantly, our SBCs can be used to simulate asymmetric lattices with one or more special boundaries.

Our paper is organized as follows.  In Section \ref{sect:methods}, we describe
how our hierarchy of SBCs can be generated by sampling the spin flip statistics
of larger and larger clusters, and how we re-sample flips of the boundary
pseudospins from these distributions. We then describe how we refresh the spin
flip statistics so that the simulation becomes self-consistent eventually, which
we check through simple measurements.  In Section \ref{sect:results}, we compare
the performances of the zeroth-order (SBC0) and first-order (SBC1) SBCs against
that of the PBC, by measuring the specific heat, magnetic susceptibility, and
spin-spin correlation function of the Ising lattice.  In Section
\ref{sect:aperiodic}, we simulate two asymmetric lattices using  SBC0 and SBC1.
In the magnetized boundary simulation, one boundary of the Ising lattice is
coupled to pseudospins which are always up, to simulate the effect of a
short-range DC magnetic field at the boundary.  In the open boundary simulation,
one boundary of the Ising lattice is left free, i.e. not coupled to any
pseudospin, to simulate an actual surface.  Along the other three boundaries, we
impose SBCs with spin flip statistics that can self-consistently vary with
distance from the special boundary in both cases.  We then conclude in Section
\ref{sect:conclusions}.

\section{Methods}
\label{sect:methods}

\subsection{Overview of Boundary Conditions}

\begin{figure}[htbp]
\centering
\includegraphics[scale=0.45]{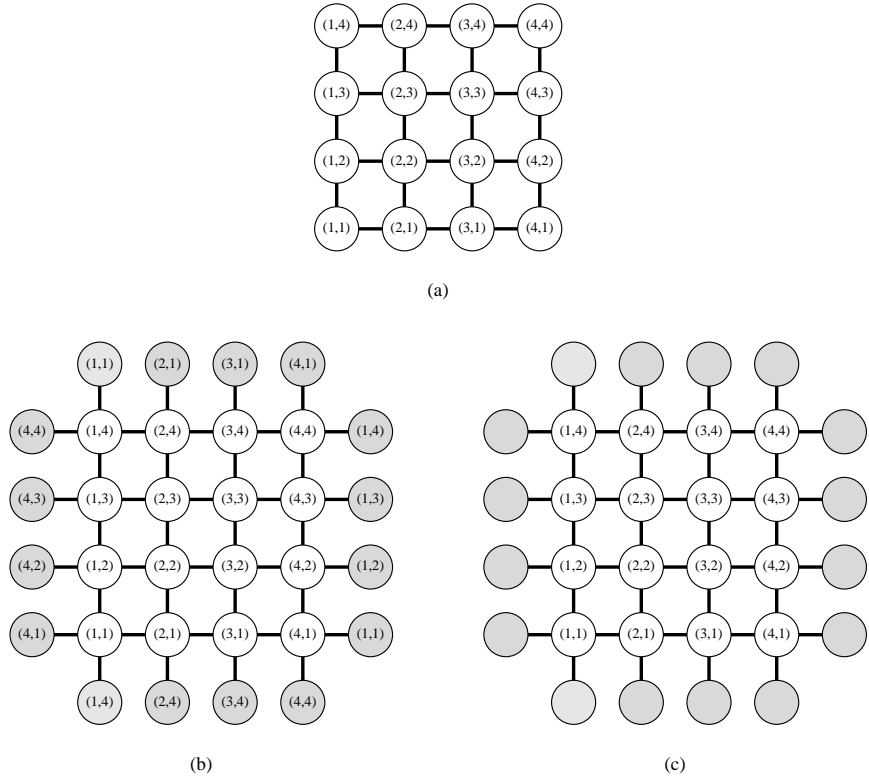}
\caption{Simulating a $4 \times 4$ Ising lattice using (a) open boundary
conditions; (b) periodic boundary conditions; and (c) stochastic boundary
conditions.  With open boundary conditions (a), the coordination numbers, three
and two respectively, of the boundary spins and corner spins are smaller than
the coordination number, four, of the interior spins.  With periodic boundary
conditions (b), all spins are coupled to four other spins, but spins on the
boundaries will be artificially correlated with spins on the opposite
boundaries.  These artificial correlations are removed in the stochastic
boundary conditions (c), where boundary spins are coupled to independent
pseudospins.} 
\label{fig:pbcsbc} 
\end{figure}

In a computer simulation, the number of spins $N$ is necessarily finite.  Unlike
the infinite Ising lattice considered in the thermodynamic limit, a finite
lattice of spins presents boundaries, which must be treated with care.  If we
choose to simulate an Ising lattice using open boundary conditions, then as
shown in Figure \ref{fig:pbcsbc}(a), different spins will have different
coordination numbers.  The translational symmetry of the infinite Ising lattice
will thus not be preserved.  In practice, PBCs are imposed, as suggested by
textbooks on Monte Carlo methods
\cite{Newman1999MonteCarloMethodsinStatisticalPhysics,
Landau2009AGuidetoMonteCarloSimulationsinStatisticalPhysics,
Binder2002MonteCarloSimulationsinStatisticalPhysics}.  With PBCs, all spins have
the same coordination number, as shown in Figure \ref{fig:pbcsbc}(b), and are
thus translationally equivalent.  However, spins on the boundaries become
artificially correlated with spins on the opposite boundaries.  When the correlation length is much shorter than the size of the simulation system, these artificial correlations do not affect measurements performed in the Monte Carlo simulation.  However, they will pose a problem when the correlation length becomes comparable to the size of the simulation system, at temperatures close to the critical temperature $T_C$.  Typically, finite size scaling analyses are
performed to extract the infinite-lattice limits of measured quantities
\cite{Binder1972Physica62p508,  Binder1981PhysRevLett47p693,
Binder1989JStatPhys55p87, Binder1992LecNotesPhys409p59,
Privman1990FiniteSizeScaling, Promberger1995ZeitsPhysB97p341,
Caracciolo1995PhysRevLett74p2969}.

If instead of PBCs, we couple each spin on the boundaries to an independent
pseudospin, as shown in Figure \ref{fig:pbcsbc}(c), it is also possible to
ensure that all system spins are coupled to the same number of nearest
neighbors.  Pseudospins do not contribute towards measurements, which are done
only over the system spins, but must also be updated from time to time, to
simulate the coupling of the system of $N$ spins to a fluctuating environment.
Our goal is to choose an appropriate pseudospin dynamics to mimic the
fluctuating environment seen by a subsystem of $N$ spins within an infinite
Ising lattice at temperature $T$.

\subsection{Hierarchy of Stochastic Boundary Conditions}

\begin{figure}[htb]
\centering
\includegraphics[scale=0.45]{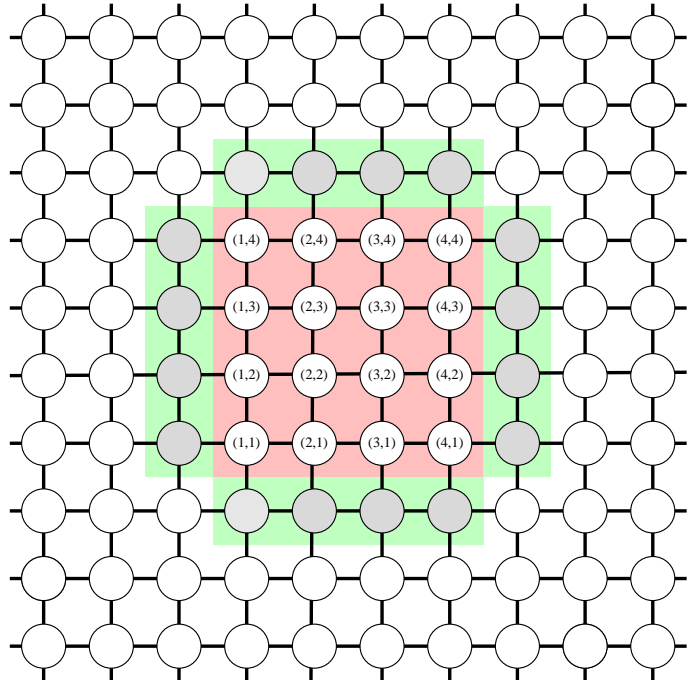}
\caption{From Monte Carlo simulations of the infinite Ising lattice, the
dynamical histories of the 16 environment spins (colored light gray) can be used
as stochastic boundary conditions for the $4 \times 4$ system they are coupled
to.} 
\label{fig:infinite} 
\end{figure}

Clearly, if we could simulate an infinite Ising lattice, we would simply extract
an ensemble of histories of the environment spins coupled to the $N$-spin
system, as shown in Figure \ref{fig:infinite}, and use these as our SBCs.  Such
a set of SBC would in fact be exact, i.e. the values of all quantities that can
be measured entirely within the $N$-spin system would be identical to their
infinite-lattice values.  This is because all possible correlations between the
$N$-spin system and its infinite environment have been incorporated into the
dynamics of the SBCs.

In practice, the SBCs could not be devised in this manner.  We should also not simulate a larger Ising lattice (with PBCs), within which the $N$-spin system is embedded, extract the dynamical histories of the relevant environment spins, and use these as our SBCs.  These SBCs capture exactly the correlations between the $N$-spin system and its finite environment, and therefore will not do better than direct simulations of the larger finite lattice in approximating the infinite-lattice limits of various observables.

However, by observing translational and rotational symmetries of the infinite
Ising lattice, we can develop a hierarchy of SBCs that approximates the
infinite-lattice embedding better and better.  First, let us observe that the
infinite Ising lattice is translationally invariant, and thus the dynamics of
the environment spins must be statistically identical to the dynamics of the
system spins.  At the zeroth order, an environment spin must flip no faster, or
no slower than a system spin.  If we call the time interval between two
consecutive flips ($\uparrow$-to-$\downarrow$ or $\downarrow$-to-$\uparrow$) of
a given spin the \emph{flip time} of the said spin, the flip time distributions
of an environment spin must be identical to that of a system spin.  Therefore,
to mimic the fluctuating environment of a $N$-spin system we design an SBC whereby the pseudospin flip statistics are identical to the system
spin flip statistics.   We call this a \emph{zeroth-order stochastic boundary
conditions}  (SBC0), because no spin-spin correlations are taken into
consideration.

In general, for the ferromagnetic Ising model, a $\downarrow$ spin surrounded by
$\uparrow$ spins is much more likely to flip compared to an $\uparrow$ spin
surrounded by $\uparrow$ spins.  In contrast to the plain spin flip statistics,
the spin flip statistics conditioned by the spin configuration of its nearest
neighbors contain information on the spin-spin correlations.  Therefore, we
would like to flip a pseudospin more frequently if it is misaligned with the
system spin it is coupled to, and less frequently otherwise.  Again, because of
translational symmetry in the Ising model, the pseudo-system spin pair should
have the same dynamics as any nearest-neighbor pairs within the system.  If we
design the SBCs such that the conditional pseudospin flip statistics are
identical to the conditional spin flip statistics, we expect to capture part of
the spin-spin correlations in the infinite Ising lattice.  We call this
a \emph{first-order stochastic boundary conditions} (SBC1), because correlations
between nearest-neighbor spins are partly accounted for.

Going further, we note that next-nearest neighbor spins make more and more
significant contributions to the correlations between spins as the correlation
length increases.  To capture this correlation, we need to go to higher-order
conditional spin flip statistics, whereby the spin configuration of next-nearest
neighbors, in addition to nearest neighbors, are taken into consideration.  Each
pseudospin has one nearest-neighbor system spin, and two next-nearest-neighbor
system spins.  In principle, the pseudospin flip statistics must be conditioned
on the configuration of these three spins, for which there are eight.  Once
again, these conditional pseudospin flip statistics should be identical to those
measured within the system.  We call this the \emph{second-order stochastic
boundary conditions} (SBC2).  We expect to capture more of the spin-spin
correlations within the infinite Ising lattice, because correlations between
next-nearest-neighbor spins are also partly accounted for.

\begin{figure}[htbp]
\centering
\includegraphics[scale=0.45]{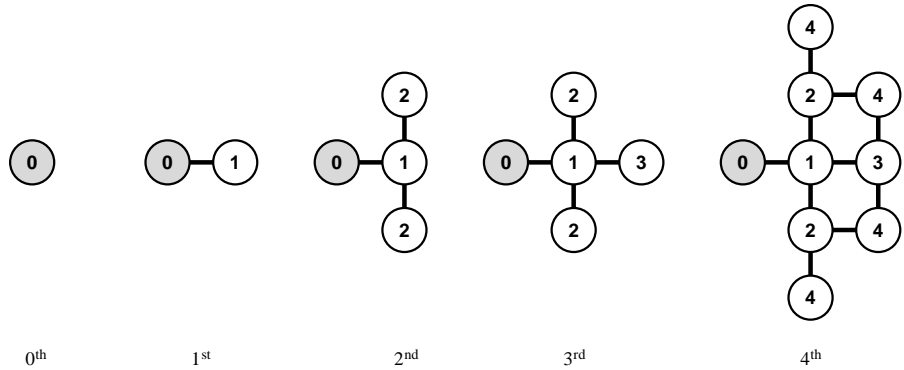}
\caption{Hierarchy of spin clusters whose conditional spin flip statistics can
be applied to stochastic boundary conditions.} 
\label{fig:hierarchy}
\end{figure}

In this way, a hierarchy of SBCs that will better and better approximate a
finite system of spins embedded within an infinite environment can be built up,
as shown in Figure \ref{fig:hierarchy}.  At some point, our measurements will so
closely approximate the infinite-system results that we do not refine the SBCs
any further.  At high temperatures, where the spin-spin correlation length is
short, we will show in Section \ref{sect:results} that we do sufficiently well with zeroth and first order SBCs.  Close to $T_C$, we will need higher-order SBCs.  In addition to finite size scaling, where simulation systems of different sizes are employed, we can also do finite order scaling, where we simulate finite
Ising lattices using SBCs of different orders, and then extrapolate to infinite
order.

\subsection{Overview of Algorithm}

Unlike for PBC, we can only start a SBC simulation after first knowing the
conditional spin flip statistics.  These can be obtained from theory, but we
prefer not to.  Instead, we will measure the spin flip statistics in a
\emph{calibration stage}, where the finite Ising lattice is simulated using PBC.
After the system equilibriates, i.e. total energy and magnetization become
stable, we start collecting flip time statistics.  In this calibration stage, we
will need a long time to accumulate enough statistics if we monitor only a
single observation cluster, which typically flips after a large number of Monte Carlo moves.  Since the infinite Ising lattice is translationally invariant, spin flip statistics from observation clusters at different sites should be identical.  Furthermore, since the Ising lattice is rotationally invariant, spin flip statistics from observation clusters at different orientations should also be identical.  We can
therefore take advantage of these symmetries to accumulate spin flip statistics
faster from multiple (and overlapping) observation clusters. 

Once the spin flip statistics have been measured, we will turn off PBC, and
impose SBC.  We call this the \emph{implementation stage}.  For a $L \times L$
square Ising lattice coupled to $4L$ pseudospins, the Hamiltonian is given by
\begin{equation} 
H = \sum_{\langle \bi, \bj \rangle} S_{\bi} S_{\bj} + 
\frac{1}{2} \sum_{\langle \bk, \bk' \rangle} S_{\bk} S_{\bk'}.  
\end{equation} 
Here, $\langle \bi, \bj \rangle = \langle (i_1, i_2), (j_1, j_2) \rangle$
denotes nearest-neighbor system sites, whereas $\langle \bk, \bk' \rangle =
\langle (k_1, k_2), (k'_1, k'_2) \rangle$ denotes nearest-neighbor
system-pseudospin sites.  $S_{\bi} = +1$ represents an $\uparrow$ spin while
$S_{\bi} = -1$ represents a $\downarrow$ spin at site $\bi$ (pseudospin site is
indicated with a prime).  The Metropolis algorithm is used to flip the system
spins, while the pseudospins are updated using the SBC algorithms described
below. 

\subsubsection{Zeroth-Order SBC}

For SBC0, we need only measure the unconditional spin flip statistics.
Monitoring the spin at $\bi$ for a long time, we will find it flipping at times
$(t_1, t_2, \dots, t_M)$.  This time series is autocorrelated, more strongly so
when the system is closer to $T_c$.  In SBC0, we will assume the spin flips to
be uncorrelated, and determine the flip times $(\Delta t_1, \dots, \Delta
t_{M-1})$, where $\Delta t_m = t_{m+1} - t_m$.  We then store the
$\uparrow$-to-$\downarrow$ flip times in queue $A$ and the
$\downarrow$-to-$\uparrow$ flip times in queue $B$.  

After enough flip time data is collected in the calibration stage, we can impose
SBC0 and start the implementation stage, by randomly initializing each
pseudospin to be $\uparrow$ or $\downarrow$.  If a pseudospin $\bk'$ is
$\uparrow$, we then randomly draw one waiting time $\Delta t$ from queue $A$.
Else, we draw $\Delta t$ from queue $B$.  We will then wait $\Delta t$ time
steps, before flipping the pseudospin $\bk'$, and draw a new waiting time
$\Delta t$ from the appropriate queue.

\begin{figure}[htbp]
\centering
\includegraphics[scale=0.375]{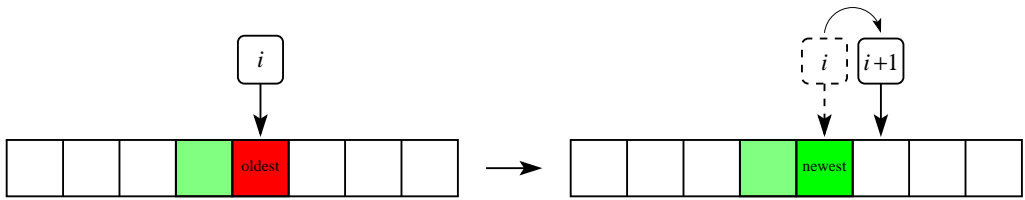}
\caption{Overwriting the oldest flip time with the newest flip time with the aid
of a moving index in a fixed-length array.} 
\label{fig:queue} 
\end{figure}

To achieve self-consistency, both queues are implemented in C as arrays with
fixed length $M$.  This data structure mimics a first-in-first-out queue for
data collection, at the same time offering us the advantage of random access for
resampling waiting times.  After the queues are filled up from the front, we
continue to acquire spin flip statistics from within the system.  The newest
flip time data that we obtain each time will overwrite the oldest flip time data
with the aid of a moving index, as shown in Figure \ref{fig:queue}.  In this
way, only the $M$ most recent flip times are stored in each queue, to ensure
that the simulation `forgets' that it started with PBC.  Measurements are taken
only after the simulation becomes self-consistent.

\subsubsection{First-Order SBC}

For SBC1, instead of single spins, we consider spin pairs and distinguish
between four different cases. For a given spin pair $S_1S_2$, we call the first spin
the \emph{target spin}, and the second spin its \emph{neighbor spin}.  We
measure the spin flip statistics of the target spin, conditioned on the state of
the neighbor spin.  To store the flip time data of the target spin, we now need
four different queues, $A_{\uparrow}$
($\uparrow\uparrow$-to-$\downarrow\uparrow$), $B_{\uparrow}$
($\downarrow\uparrow$-to-$\uparrow\uparrow$), $A_{\downarrow}$
($\uparrow\downarrow$-to-$\downarrow\downarrow$), and $B_{\downarrow}$
($\downarrow\downarrow$-to-$\uparrow\downarrow$).  Because there are now four
instead of two queues to fill, we must collect more spin flip statistics from the simulation.  Therefore, whenever a system spin is flipped, we use all four spin pairs it is part of to update the queues.

\begin{figure}[htb]
\centering
\includegraphics[scale=0.375]{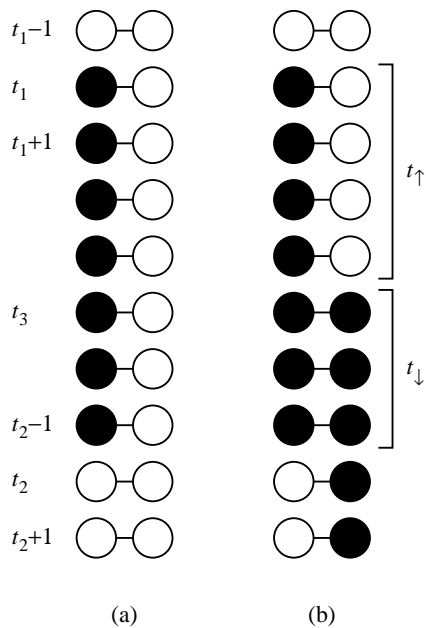}
\caption{Measuring the flip time of the left target spin when (a) the right
neighbor spin stays spin-$\uparrow$ between $t_1$ and $t_2$; and (b) when the
right neighbor spin flips at some $t_1<t_3<t_2$.}
\label{fig:sbc1}
\end{figure}

To understand how the conditional spin flip statistics is collected for SBC1,
consider the two examples shown in Figure \ref{fig:sbc1}.  In Figure
\ref{fig:sbc1}(a), the left target spin flips from $\uparrow$ to $\downarrow$ at
time $t_1$, and then back to $\uparrow$ at time $t_2$.  Throughout this time
interval, the neighbor spin remains $\uparrow$.  Therefore, we store the flip
time $\Delta t = t_2 - t_1$ in the queue $B_{\uparrow}$.  In Figure
\ref{fig:sbc1}(b), the neighbor spin flips from $\uparrow$ to $\downarrow$ at
time $t_1 < t_3 < t_2$.  Therefore, the $\downarrow$-to-$\uparrow$ spin flip of
the target spin at $t_2$ is conditioned $t_{\uparrow} = t_3 - t_1$ of the time
by a $\uparrow$ neighbor spin, and $t_{\downarrow} = t_2 - t_3$ of the time by a
$\downarrow$ neighbor spin.  We add the flip time $\Delta t = t_2 - t_1$ to two
queues, $B_{\uparrow}$ with weight $w_{\uparrow} = t_{\uparrow}/\Delta t$, and
$B_{\downarrow}$ with weight $w_{\downarrow} = t_{\downarrow}/\Delta t$.  We can
also understand the data collection for Figure \ref{fig:sbc1}(a) in terms of
these weights: since $t_{\downarrow} = 0$, $w_{\uparrow} = 1$ and
$w_{\downarrow} = 0$.

As with SBC0, pseudospins will be randomly initialized at the start of the
implementation stage.  There are then two ways to draw a waiting time for a
pseduospin.  In the first approach, we first draw a continuous random number $W$
from the uniform distribution $U(0, W_M)$, where $W_M = \sum_{t=1}^M w_t$ is the total weight stored in the flip time array.  We then run through the weight array until
the cumulative weight $\sum_{t=1}^{t^*} w_t \geq W$.  The $t^*$th entry in the
flip time array will be our new waiting time.  Alternatively, we can start by
selecting a random entry in the flip time array, whose weight is $w$.  We then
draw a continuous random number $r$ from $U(0, 1)$.  If $r < w$, we accept this
entry as our new waiting time.  Else, we will select a new random entry and a
new $U(0,1)$ random number $r$, until the random entry is accepted.  In both
approaches, the probability of a waiting time being selected is proportional to
its weight in the flip time array.  The first approach uses the random number
generator efficiently, because every trial is accepted, but is slow because of
the need to sum the weights.  The second approach wastes some random numbers,
because not every trial is accepted, but is fast because no cumulative sums are
evaluated.  In this study, we adopted the second approach for generating waiting
times.

More importantly, while we are waiting for a $\uparrow$ pseudospin to flip, its
neighbor system spin may flip from $\uparrow$ to $\downarrow$, or vice versa.
If the neighbor system spin had remained $\uparrow$ ($\downarrow$) throughout,
we should sample the pseudospin waiting time $\Delta t_{\uparrow\uparrow}$
($\Delta t_{\uparrow\downarrow}$) from $A_{\uparrow}$ ($A_{\downarrow}$).  When
the neighbor system spin flips back and forth between $\uparrow$ and
$\downarrow$, the proper waiting time $\Delta t_{\uparrow}$ should be a linear
combination of $\Delta t_{\uparrow\uparrow}$ and $\Delta
t_{\uparrow\downarrow}$.  We also expect the contribution from $\Delta
t_{\uparrow\uparrow}$ ($\Delta t_{\uparrow\downarrow}$) to be proportional to the
total time the neighbor system spin is $\uparrow$ ($\downarrow$) during the
waiting time.  However, since we do not know when the neighbor system spin will
flip, we cannot calculate the waiting time $\Delta t_{\uparrow}$ right after the
pseudospin flipped.

To solve this problem, we introduce a \emph{waiting fraction} $f_w$, which is the fraction of total waiting time completed by a pseudospin.  Right after a pseudospin flipped, we draw $\Delta t_{\uparrow\uparrow}$ and $\Delta t_{\uparrow\downarrow}$ from $A_{\uparrow}$ and $A_{\downarrow}$, and set $f_w = 0$.  For subsequent time steps, if the neighbor system spin is $\uparrow$, $f_w$ increases by $1/\Delta t_{\uparrow\uparrow}$, whereas if the neighbor system spin is $\downarrow$, $f_w$ will increase by $1/\Delta t_{\uparrow\downarrow}$.  In this way, $f_w$ will increase at an average rate
determined by the proportions of times the neighbor system spin spends in the
$\uparrow$ and $\downarrow$ states.  When $f_w$ reaches or exceeds $1$, we flip the
pseudospin.

\subsection{Self-Consistency}

At the start of the SBC simulations, the boundary conditions are inconsistent,
because we are flipping pseudospins using spin flip statistics obtained from
PBC.  These would be contaminated with artificial correlations.  Therefore,
while the SBC simulations are running, we keep measuring the spin flip
statistics, and use the updated spin flip statistics for the SBC simulations.
Because spins on opposite boundaries are no longer coupled to each other,
artificial PBC correlations will die out, and we will be left with whatever
approximation of the infinite-lattice correlations the SBC admits.  Eventually,
the spin flip statistics will no longer change with further updating, and we say
that the SBC simulation has become \emph{self-consistent}.  Measurements can
then begin.

To check that artificial PBC correlations in the flip time statistics collected
from the calibration stage indeed die out in the implementation stage, and how
long it takes for the system to become self-consistent, we examine the
distributions $K(t)$ of flip times in the data queues.  We measure changes in
these distributions in two ways, by (1) computing the Jensen-Shannon divergence
(JSD) between the current distribution and the distribution at an earlier time,
as well as (2) by measuring the moments of the old and new distributions.

\subsection{Measurements}

Once the SBC simulation is self-consistent, we start the measurements of various
physical quantities.  We do this in the high temperature regime, where low order
SBCs are expected to be able to capture most of the weak correlations between
spins, for three scenarios:
\begin{enumerate}
\item an $L \times L$ Ising lattice embedded within an infinite system.  Such
simulations are labeled SBC0 and SBC1 depending on whether the zeroth-order or
first-order algorithm has been used;
\item an $L \times L$ Ising lattice embedded within a semi-infinite system, such
that the pseudospins along one boundary are perfectly magnetized.  The other
three boundaries are coupled to a fluctuating environment.  Such simulations are
labeled SBC0 + M and SBC1 + M depending on which SBC algorithm has been used;
and
\item an $L \times L$ Ising lattice embedded within a semi-infinite system, such
that one boundary is completely open (i.e. not coupled to pseudospins).  The
other three boundaries are coupled to a fluctuating environment.  Such
simulations are labeled SBC0 + O and SBC1 + O depending on which SBC algorithm
has been used.
\end{enumerate}

The purpose of the first scenario is to compare the performances of the SBC
simulations against the PBC simulation, as well as against analytical results.
We simulate the second and third scenarios because these cannot be easily done
using PBC, to demonstrate the versatility of SBCs.

\subsubsection{Correlation Time}
\label{corrtime}

The first quantity we measure in each simulation is the correlation time $\tau$.
In each Monte Carlo time step, at most one spin is flipped and physical
properties of the system remain largely the same.  Therefore the data collected
for consecutive time steps will be highly correlated.  To obtain statistically
independent data points, we need to let the system evolve for a time on the
order of $\tau$.

To measure $\tau$, we compute the autocorrelation function 
\begin{equation}
\chi(t)=\frac{1}{t_{\max}-t} \sum_{t'=1}^{t_{\max}-t} m(t')m(t'+t) - \langle m \rangle^2
\end{equation}
of the average magnetization $m$, and fit it to a decaying exponential of the
form
\begin{equation}
\chi(t) = \chi(0)\exp\left(-\frac{t}{\tau}\right).
\end{equation}
For the rest of the physical quantities, we then perform statistically
independent measurements every $2\tau$
\cite{Newman1999MonteCarloMethodsinStatisticalPhysics}. 

\subsubsection{Other Measured Quantities}

For the first scenario, where our goal is to compare SBC simulations against PBC
simulations, we measure the specific heat 
\begin{equation}
c(T) = \beta^2 N \left(\langle e^2 \rangle-\langle e \rangle^2 \right),
\end{equation}
and the magnetic susceptibility
\begin{equation}
\chi(T) = \beta N \left( \langle m^2 \rangle-\langle m \rangle^2 \right),
\end{equation}
which are defined in terms of the variances of the average energy per spin
\begin{equation}
e = \frac{1}{N}\left(\sum_{\langle \bi, \bj \rangle} S_{\bi} \cdot S_{\bj} + 
\frac{1}{2} \sum_{\langle \bk, \bk' \rangle} S_{\bk} \cdot S_{\bk'}\right),
\end{equation}
and the average magnetization per spin
\begin{equation}
m = \frac{1}{N}\sum_{\bi} S_{\bi},
\end{equation}
Here, $\beta = 1/k_B T$, with the Boltzmann constant $k_B = 1$ for our
simulations, $T$ is the temperature, and $N$ is the total number of
system spins.  The notation $\langle \bi, \bj \rangle$ indicates that $\bi$ and
$\bj$ are nearest neighbors while $\langle \bk, \bk' \rangle$ indicates the
pseudospin at $\bk'$ is a nearest neighbor of the spin at $\bk$.  Finally,
$\langle \cdot \rangle$ indicates an ensemble average over statistically
independent samples from multiple simulations,

We also measure the spin-spin correlation 
\begin{equation}
G(\bi, \bj) = \langle S_{\bi} S_{\bj} \rangle - \langle S_{\bi} \rangle \langle
S_{\bj} \rangle.
\end{equation}
between two spins at $\bi$ and $\bj$.  Since the Ising lattice is
translationally and rotationally invariant, we can average $G(\bi, \bj)$ over
all pairs $(\bi, \bj)$ with the same separation $r = |\bi - \bj|$ to get
\begin{equation}
G(r) = \frac{1}{n(r)} \sum_{|\bi - \bj| = r} G(\bi, \bj),
\end{equation}
where $n(r)$ is the number of pairs with separation $r$.

\subsubsection{Measurements for Magnetized and Open Boundaries}

For the magnetized and open boundaries, we would like to see how a \emph{single}
special boundary affect physical quantities like the magnetization per spin and
magnetic susceptibility at different distances from the boundaries.  To do this,
we measure the magnetization per spin 
\begin{equation}
m(i) = \frac{1}{L} \sum_{j=1}^L S_{(i,j)}
\end{equation}
$i$ rows away from the special boundary.  The average magnetization and magnetic
susceptibility $i$ rows away from the special boundary are then $\langle m(i)
\rangle$ and
\begin{equation}
\chi(i) = \beta L \left(\langle m(i)^2 \rangle-\langle m(i) \rangle^2\right).
\end{equation}
Here $L$ is the number of spins in one row. We do this one row at a time because
the presence of a special boundary breaks the translational invariance in the
direction perpendicular to the boundary, but translational invariance is
maintained in the other direction.

We also expect the spin-spin correlation to vary from row to row because of the
special boundary. More importantly, rotational invariance is also broken by the
presence of the special boundary, so we have two distinct limits for the
spin-spin correlation function.  We call correlations between spins in the
same row (same distance from the special boundary) the \emph{transverse spin-spin
correlation} $G_{\perp}(r)$ and correlations between spins in the same column
(different distances from the special boundary) the \emph{longitudinal spin-spin
correlation} $G_{\parallel}(r)$.

\section{Performance of SBC}
\label{sect:results} 

\subsection{Self-Consistency}

For SBC0 simulations, each datum has the same weight. If queue $A$ has $M$
entries, and the flip time $t$ appears $n(t)$ times, then $K(t) = n(t)/M$.  For
SBC1 simulations, each datum has a different weight. The total weight of the
data in queue $A_{\uparrow}$ is $W_{\uparrow} = \sum_{i=1}^{M} w_{\uparrow}(i)$,
where $w_{\uparrow}(i)$ is the weight of the $i$th flip time stored in the queue
$A_{\uparrow}$. If the flip time $t$ occurs in entries {$t_1$, $t_2$, \dots
$t_{n(t)}$}, the probability $K_{\uparrow}(t)$ for finding the target spin flip
from $\uparrow$ to $\downarrow$ in $t$ time steps given its neighbor spin is
$\uparrow$ will be given by 
\begin{equation}
K_{\uparrow}(t)=\frac{1}{W_{\uparrow}} \sum_{i=1}^{n(t)} w_{\uparrow}(t_i).
\end{equation}

To test for self-consistency, we measure the distributions $K_n(t)$ at a set of
discrete times $\{\tau_n\}$.  The time interval $\Delta\tau = \tau_{n+1} -
\tau_n$ is chosen to be large enough for a significant fraction of the queues to
be refreshed, but small enough that we can still monitor the long-time evolution
of $K_n(t)$.

\subsubsection{Jensen-Shannon Divergence}

For two flip time distributions $K_1(t)$ and $K_2(t)$, where $t = 1, 2, \dots$,
we can define their Jensen-Shannon divergence (JSD) to be
\begin{equation}
\mathrm{JSD}(K_1, K_2) = -\sum_{t=1}^{\infty} K(t)\ln K(t) +
\frac{1}{2}\sum_{t=1}^{\infty} K_1(t) \ln K_1(t) +
\frac{1}{2}\sum_{t=1}^{\infty} K_2(t) \ln K_2(t).
\end{equation}
Here,
\begin{equation}
K(t) = \frac{1}{2}\left[K_1(t) + K_2(t)\right]
\end{equation}
is the average flip time distribution obtained by combining the flip time
statistics for $K_1(t)$ and $K_2(t)$.  The Jensen-Shannon divergence \cite{Lin1991IEEETransInforTheor37p145}, which is a symmetrized version of the
Kullback-Liebler divergence \cite{Kullback1951AnnMathStats22p79, Kullback1968InforTheorStats}, measures the statistical
distance between two probability distributions.  The larger the JSD, the more
different the two distributions are.

\begin{figure}[H]
\centering
\includegraphics[scale=0.45]{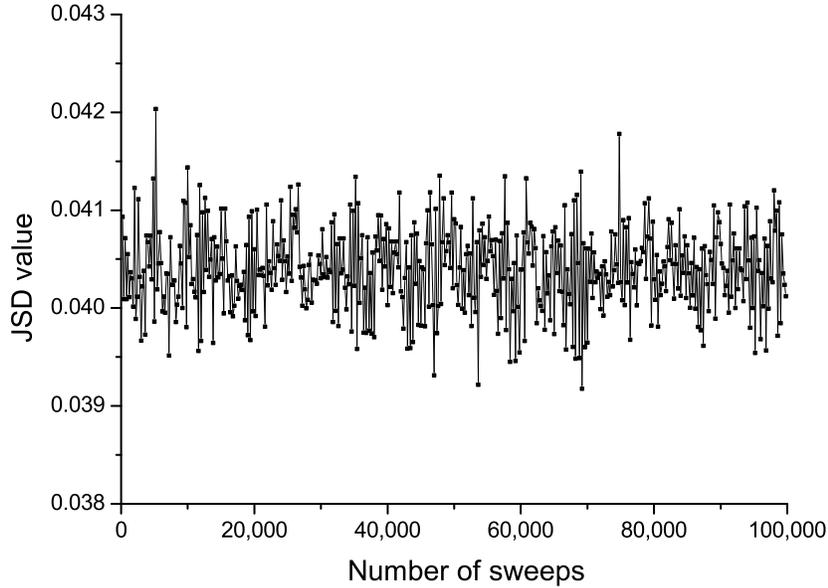}
\caption{The JSD value for the $\uparrow$-to-$\downarrow$ flip time
distributions versus the number of sweeps of a SBC0 simulation of a $100 \times
100$ Ising lattice at $T = 3.0$. The length of the queue used to collect the
$\uparrow$-to-$\downarrow$ flip time data is $M = 1,000,000$, and the time
interval between successive distributions is $\Delta\tau = 200$ sweeps.  Each
sweep consists of 10,000 Monte Carlo steps.}
\label{fig:JSD}
\end{figure}

In our study, we calculate JSD of successive flip time distributions $K_n(t)$
and $K_{n+1}(t)$ to observe possible changes in the distributions.  We expect
\begin{equation}
\mathrm{JSD}_n = \mathrm{JSD}(K_n, K_{n+1})
\end{equation}
to decrease initially, and fluctuate about a constant value after the flip time
distribution has converged.  However, when we plot the JSD value of the
$\uparrow$-to-$\downarrow$ flip time distributions for a SBC0 simulation of a
$100 \times 100$ Ising lattice at $T = 3.0$ as a function of time in Figure
\ref{fig:JSD}, we see that the JSD value fluctuates weakly about an average
value of JSD = 0.0404.  To better appreciate this average JSD, let us note that the numerical value of the JSD represents the difference between two distributions in number of bits.  There is thus about 0.04 bits of difference, or $4 \times 10^{-8}$ bits of difference per data points, between successive flip time distributions, even
though a significant fraction of the queue has been refreshed in the intervening
time.  In particular, this suggests that the flip time distribution estimated
from the PBC calibration stage is already very close to the self-consistent flip
time distribution.  To confirm this, we also calculated the moments of the
distributions.

\subsubsection{Moments of flip time distributions}

\begin{figure}[H]
\centering
\includegraphics[scale=0.45]{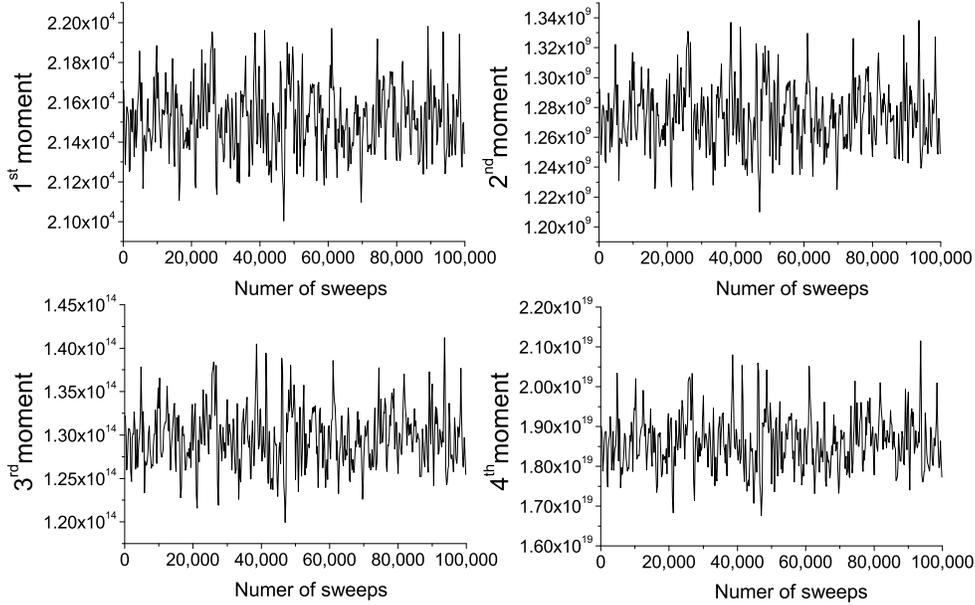}
\caption{The first to fourth moments of the $\uparrow$-to-$\downarrow$ flip time
distributions in the SBC0 simulation at $T = 3.0$ for a $100 \times 100$ square
Ising lattice, versus the number of 10,000-Monte-Carlo-step sweeps.}
\label{fig:moment1}
\end{figure}

The $k$th moment $\mu_{k}$ of a flip time distribution $K(t)$ is
\begin{equation}
\mu_{k} = \sum_t t^k K(t).
\end{equation}
Any change in $K(t)$ would therefore be reflected as changes in its moments.
Inspecting the first four moments for the $\uparrow$-to-$\downarrow$ flip time
distribution at $T = 3.0$ for a $100 \times 100$ lattice in Figure
\ref{fig:moment1}.  As we can see, these are just fluctuating about constant
average values.  Therefore, as with the JSD calculation, we conclude that the
flip time distribution obtained during the PBC calibration stage is already very
close to the steady-state flip time distribution, so our simulations became
self-consistent very shortly after SBC is turned on.

\subsection{Comparison Against PBC}

\begin{figure}[htbp]
\centering
\includegraphics[scale=0.35]{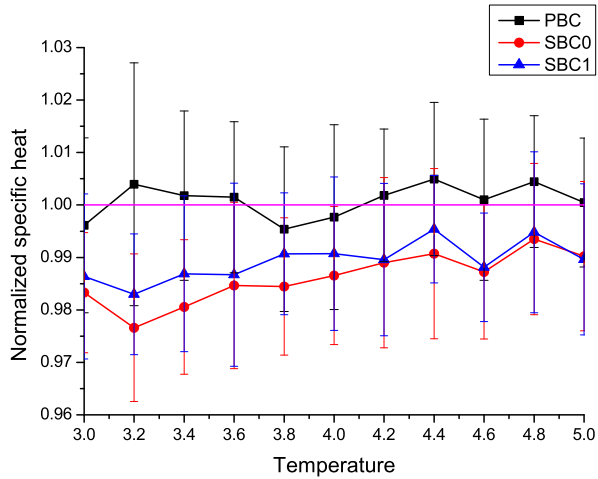}
\caption{Normalized specific heats measured from PBC, SBC0 and SBC1
simulations of a $100 \times 100$ square Ising lattice.  We ran 10 simulations
for each boundary condition and each temperature.  For each simulation,
the specific heat $c$ is calculated from 10,000 statistically independent
values of the average energy per spin $e$ sampled every $2\tau$.  The error bar
for $c$ is then the standard deviation over the 10 equivalent simulations.}
\label{fig:norsh}
\end{figure}

To compare SBC0 and SBC1 against PBC, we simulate each boundary condition and
each temperature ten times.  For each simulation, the specific heat and magnetic susceptibility were each calculated from 10,000 independent data points.  Since we have the analytical expression for the specific heat of a square Ising lattice
\cite{Onsager1944PhysRev65p117}, we divided the numerical specific heats by the analytical specific heat, and plot the ratio in Figure  \ref{fig:norsh}.  As we can see, the SBC0 and SBC1 specific heats are slightly smaller than the analytic specific heat, while the PBC specific heat fluctuates about the analytical specific heat.  Comparing the two SBCs, we find the SBC1 specific heat closer to the analytic specific heat.  However, these differences are not strongly significant in statistical terms.

\begin{figure}[htbp]
\centering
\includegraphics[scale=0.35]{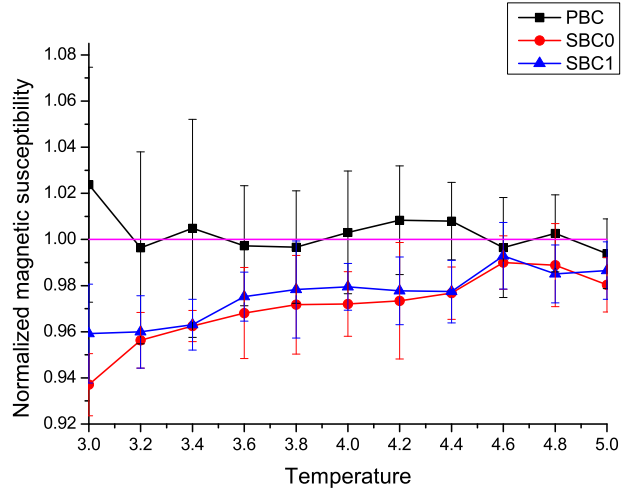}
\caption{Normalized magnetic susceptibilities measured from PBC, SBC0 and SBC1
simulations of a $100 \times 100$ square Ising lattice.  We ran 10 simulations
for each boundary condition and each temperature.  For each simulation, the
magnetic susceptibility $\chi$ is calculated from 10,000 statistically
independent values of the average magnetization per spin $m$ sampled every
$2\tau$.  The error bar for $\chi$ is then the standard deviation over the 10
equivalent simulations.}
\label{fig:ms}
\end{figure}
 
\begin{figure}[htbp]
\centering
\includegraphics[scale=0.35]{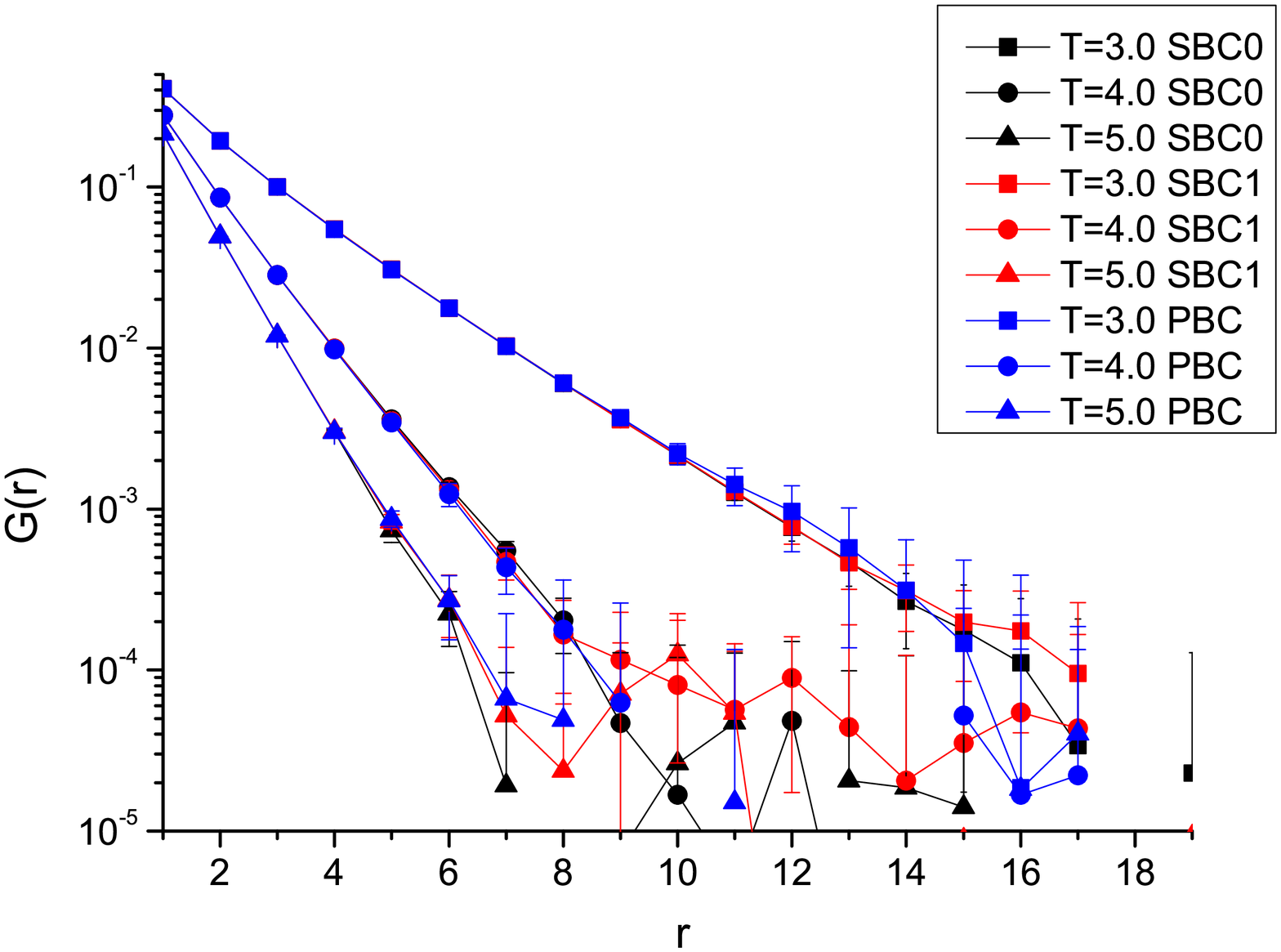}
\caption{Spin-spin correlations measured from PBC, SBC0 and SBC1 simulations of
a $100 \times 100$ square Ising lattice at $T = 3.0, 4.0, 5.0$.  For each
boundary condition and each temperature, we repeat the simulation ten times to
estimate the error bars.}
\label{fig:sscorr}
\end{figure}

In Figure \ref{fig:ms}, we show the magnetic susceptibilities $\chi$ obtained
for the three boundary conditions, divided by the magnetic susceptibility obtained using the high-temperature series expansion \cite{Boukraa2008ExpMathMagSuscSquareIsingLatt}.  Here we find that the SBC0 and SBC1 magnetic susceptibilities are both lower than the high-temperature series expansion, while the PBC magnetic susceptibility is slightly higher.  As with the specific heat, the SBC1 magnetic susceptibility is closer to the series expansion magnetic susceptibility.  None of these differences are statistically significant.  Finally, we show in Figure \ref{fig:sscorr} the spin-spin correlation $G(r)$ obtained at three different temperatures for the three boundary conditions.  For all separations, $G(r)$ for the three boundary conditions are identical to each other to within numerical uncertainties.

Through these measurements, we see that the SBC is just as good as the PBC for
simulating the two-dimensional square Ising lattice in the high-temperature
regime.  In the next section, we will illustrate the advantages SBC have over
PBC for Ising simulations, by considering Ising lattices with one special
boundary.

\section{Asymmetric Lattices}
\label{sect:aperiodic}

One very powerful application of SBC is that it can be used to simulate
asymmetric lattices, for example, the Ising lattice in a spatially varying
magnetic field.  In these lattices, translational symmetry is broken, and PBC
can no longer be imposed on all four boundaries.  However, so long as the local
flip time statistics do not vary too quickly from point to point in the Ising
lattice, we can still flip pseudospins using the flip time statistics of systems
spins on the boundary.

In this paper, we investigated the Ising lattice with one boundary coupled to
static $\uparrow$ spins, as well as one with one open boundary.  We call these
the magnetized boundary (M) and open boundary (O) respectively.  For both
lattices, we expect the influence of the special boundary to be strong close to
it, and weak further from it.  Because spins in the same row are equidistant
from the special boundary, we expect flip time statistics to be identical within
a row, and slowly varying from row to row.  This means that instead of two flip
time distributions for SBC0 or four flip time distributions for SBC1, we will
need to work with a larger number of flip time distributions, at different
distances from the special boundary.

\begin{figure}[htbp]
\centering
\includegraphics[width=\linewidth]{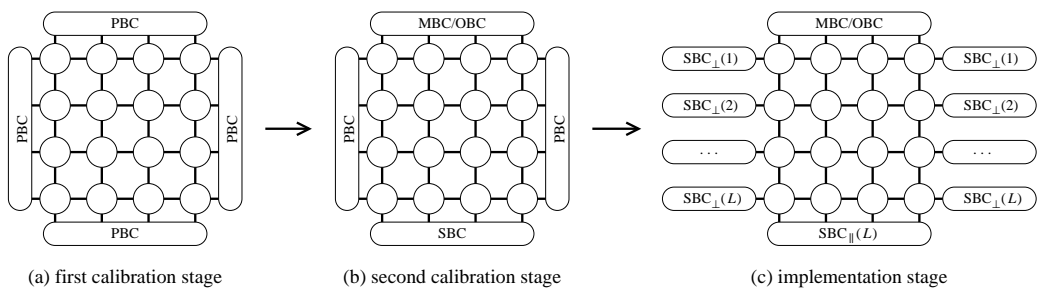}
\caption{To simulate an asymmetric Ising lattice with magnetized or open boundary conditions, we need to run the first calibration stage with PBC on all boundaries.  Once we collect sufficient spin flip statistics, we switch over to the second calibration stage, where we impose the magnetized/open boundary conditions on one boundary, SBC on the opposite boundary, and PBC on the remaining two boundaries.  In this second calibration stage, we collect spin flip statistics row by row, and also distinguish between spin pairs parallel or perpendicular to the special boundary.  In the implementation stage, we turn on SBCs that vary with distance away from the special boundary.}
\label{fig:asymlat}
\end{figure}

\subsection{Algorithm}

Two calibration stages are needed for asymmetric lattices, as shown in Figure \ref{fig:asymlat}.  In the \emph{first calibration stage}, we collect the system-wide spin flip statistics within the lattice with PBC imposed on all four boundaries.  After the system-wide spin flip statistics is obtained, we impose magnetized/open boundary conditions to one boundary and SBC to the opposite boundary.  We continue to impose PBC on the other two boundaries.  In this \emph{second calibration stage}, we collect spin flip statistics row by row, to allow them to vary with distance from the special boundary.  In this calibration stage, we also distinguish between spin pairs parallel or perpendicular to the special boundary, for higher-order SBCs.  Because we now need to construct many more flip time distributions, we use many queues containing 100,000 flip times instead of one queue containing 1,000,000 flip times.

Finally, in the \emph{implementation stage}, we implement SBC$_{\perp}$(1) one row from the special boundary, SBC$_{\perp}$(2) two rows from the special boundary, and so on and so forth till SBC$_{\perp}(L)$ on the row of system spins furthest from the special boundary.  For the row of pseudospins along the opposite boundary, we impose SBC$_{\parallel}(L)$.  We allow the simulation to become self-consistent before performing measurements.

\begin{figure}[H]
\centering
\includegraphics*[ width = 0.7\textwidth]{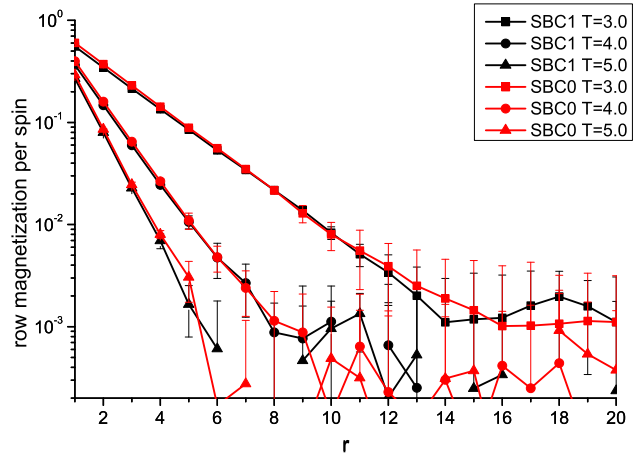}
\caption{Row magnetization per spin $m(i)$ as a function of the distance $r$ from the magnetized boundary, for a $100 \times 100$ Ising lattice with a magnetized boundary simulated using SBC0 and SBC1 at $T = 3.0, 4.0, 5.0$.}
\label{fig:mmag}
\end{figure}

\subsection{Comparison Between Orders}

For the magnetized boundary and open boundary simulations, there are no analytical results to compare against.  Therefore, we first compare the results for SBC1 against that of SBC0 for the Ising lattice with a magnetized boundary.  In Figure \ref{fig:mmag} we show the row magnetization per spin, measured using SBC0 and SBC1. As we can see, the results of the two SBCs agree very well with each other.

\begin{figure}[htb]
\centering
\includegraphics[scale=0.35]{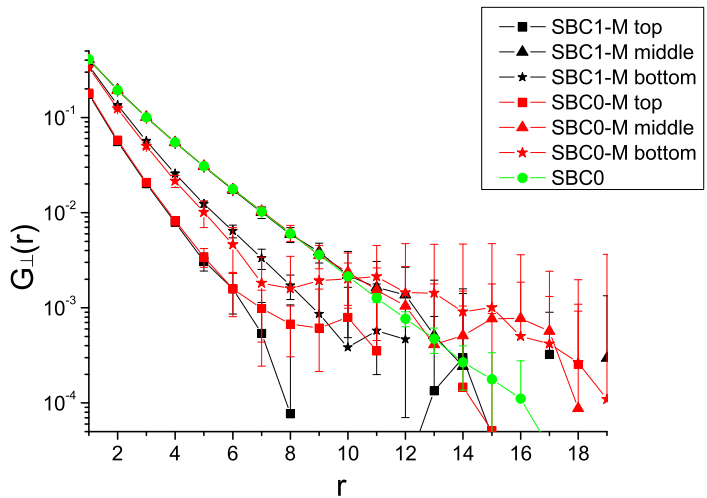}
\caption{The transverse spin-spin correlation $G_{\perp}(r)$ in the top, middle and bottom rows for a $100 \times 100$ Ising lattice with a magnetized boundary above the top row.  In this figure, we also show $G_{\perp}(r)$ for the symmetric lattice with no special boundaries.  For $r > 6$ our results are not reliable, because we used short queues containing only 100,000 flip times to implement the SBCs, as opposed to 1,000,000 flip times in the queue for the symmetric lattice.}
\label{fig:magsslrallk}
\end{figure}

We also compare the transverse spin-spin correlation function measured using SBC0 and SBC1 for an Ising lattice with a magnetized boundary, and also against the SBC0 spin-spin correlation function for a symmetric lattice with no special boundaries.  This is shown in Figure \ref{fig:magsslrallk}.  As we can see, $G_{\perp}(r)$ for SBC1 agrees very well with that for SBC0.  We also see that $G_{\perp}(r)$ in the middle of the Ising lattice with a magnetized boundary coincides with $G(r)$ of the symmetric Ising lattice with no special boundaries.  This is expected, since the middle of the asymmetric lattice is far away from the magnetized boundary.

More interestingly, we find that $G_{\perp}(r)$ is suppressed in the row adjacent to the magnetized boundary, and also the row adjacent to the opposite stochastic boundary.  Suppression of $G_{\perp}(r)$ right next to the magnetized boundary is expected, because we need to subtract from $\langle S_{\mathbf{i}} S_{\mathbf{j}} \rangle$  the product $\langle S_{\bi} \rangle \langle S_{\bj} \rangle$ (which is larger close to the magnetized boundary).  The suppression of $G_{\perp}(r)$ in the row adjacent to the stochastic boundary gives us a sense of how well the SBCs are working.  In principle, if we use an SBC with a high enough order, $G_{\perp}(r)$ in this row would be equal to $G_{\perp}(r)$ measured in the middle of the lattice, i.e. as if no boundaries were present. Indeed, the SBC1 $G_{\perp}(r)$ for this row is closer to the bulk $G_{\perp}(r)$ than the SBC0 $G_{\perp}(r)$, giving us confidence that this convergence is indeed happening.

For other physical quantities, the SBC0 values also agree very well with their SBC1 counterparts.  Therefore, from this point on, we use only the SBC1 results to compare the effects of different boundary conditions.

\begin{figure}[H]
\centering
\includegraphics*[ width = 0.7\textwidth]{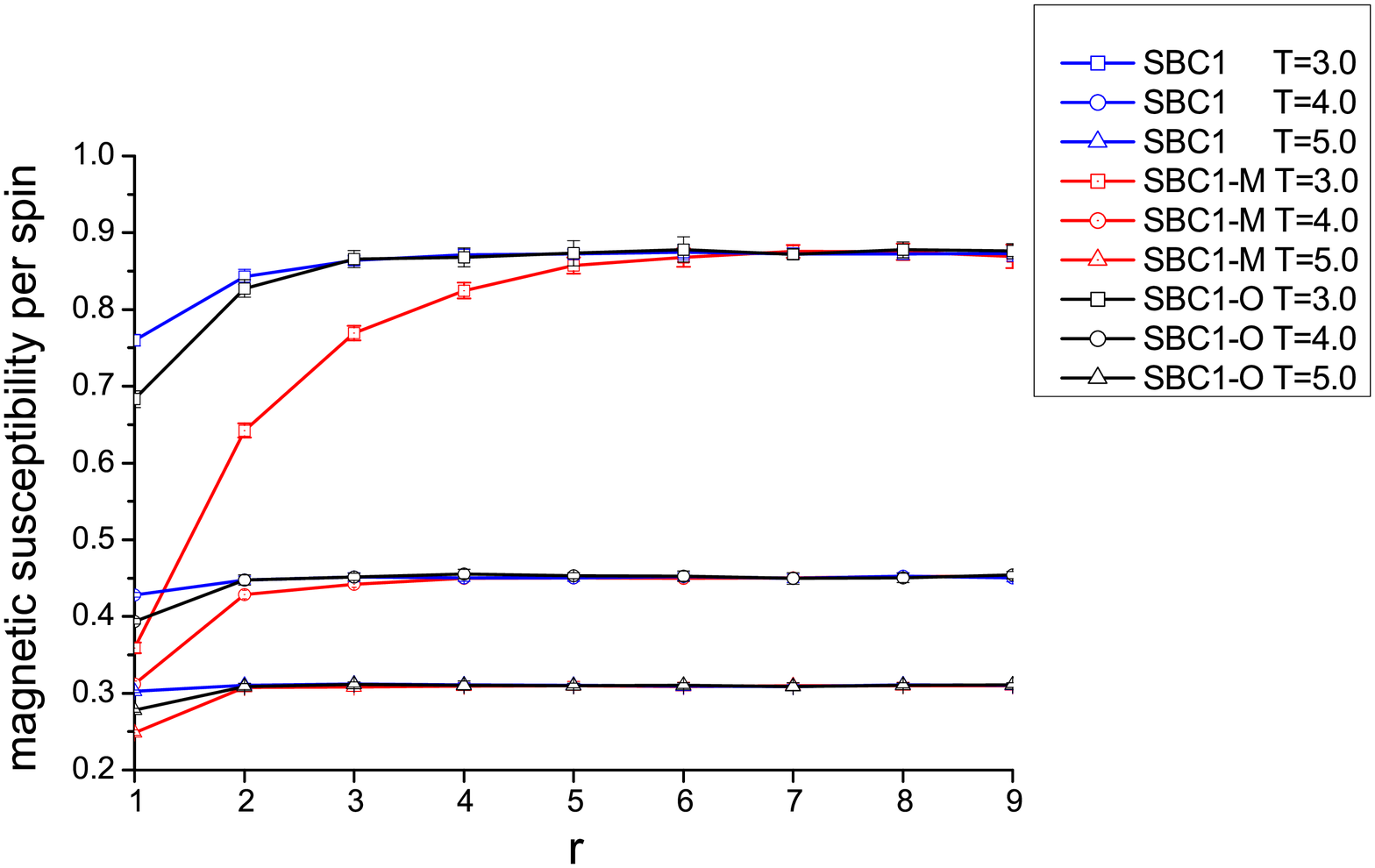}
\caption{Row magnetic susceptibility per spin as a function of distance from the boundary, measured using SBC1 from a $100 \times 100$ Ising lattice at with no special boundaries (blue), a magnetized boundary (red), and a open boundary (black) at $T = 3.0, 4.0, 5.0$.}
\label{fig:mms}
\end{figure}

\subsection{Comparison Between Boundary Conditions}

In Figure \ref{fig:mms} we compare the row magnetic susceptibilities per spin at different distances from the magnetized and open boundaries, measured using SBC1, and also compared against the row magnetic susceptibility per spin at different distances from a given boundary for the SBC1 symmetric Ising lattice.  As we can see, the row magnetic susceptibilities per spin for all three boundary conditions reach the same bulk value by a distance $r = 9$ from the boundary, for $T = 3.0, 4.0, 5.0$.  Close to the boundary, we find the row magnetic susceptibility per spin of the SBC1 symmetric lattice dip below the bulk value by about 10\%.  Since we expect an infinite-order SBC to completely eliminate the presence of a boundary, this gives a measure how `imperfect' SBC1 is.

For the Ising lattice with an open boundary, the row magnetic susceptibility per spin is suppressed to a level below the symmetric lattice.  Based on the error bars found in our measurements, this suppression of the magnetic susceptibility by the open boundary is statistically significant.  In particular, for $T = 5.0$, where SBC1 is effectively as good as SBC-$\infty$, we find no visible suppression of the row magnetic susceptibility per spin at the boundary of the symmetric lattice.  The row magnetic susceptibility per spin of the asymmetric lattice with an open boundary, on the other hand, is clearly suppressed at the boundary.  The strongest suppression of the row magnetic susceptibility per spin occurs for the asymmetric lattice with a magnetized boundary.  This suppression is especially strong at $T = 3.0$, which is very close to the critical temperature $T_C = 2.269$ of the square Ising lattice \cite{Newman1999MonteCarloMethodsinStatisticalPhysics}.

\begin{figure}[H]
\centering
\includegraphics[scale=0.35]{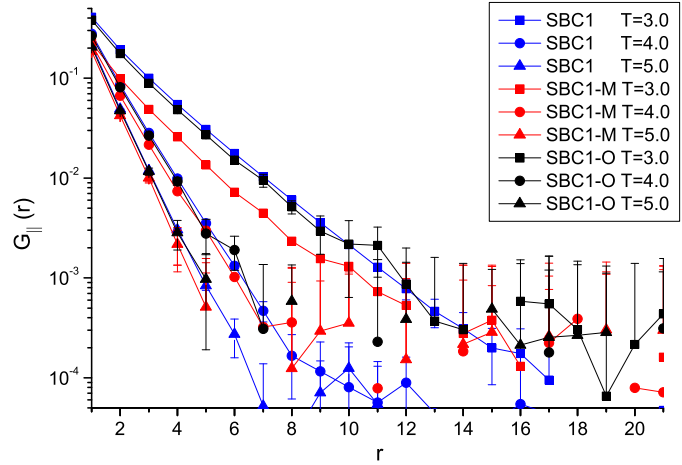}
\caption{Longitudinal spin-spin correlation $G_{\parallel}(r)$ between spins in the row adjacent to the boundary, and spins $r+1$ rows away for a $100 \times 100$ Ising lattice at $T = 3.0, 4.0, 5.0$.  In this figure, $G_{\parallel}(r)$ of the symmetric lattice with no special boundaries are shown in blue, $G_{\parallel}(r)$ of the asymmetric lattice with a magnetized boundary are shown in red, while $G_{\parallel}(r)$ of the asymmetric lattice with an open boundary are shown in black.}
\label{fig:magudallk}
\end{figure}

Next, we show in Figure \ref{fig:magudallk} the longitudinal spin-spin correlation $G_{\parallel}(r)$ between spins in the row adjacent to the boundary, and spins at various distances from the boundary.  Here we see that at higher temperatures ($T = 4.0$ and $T = 5.0$), $G_{\parallel}(r)$ coincides for all three boundary conditions.  At $T = 3.0$, which is closest to $T_C$, $G_{\parallel}(r)$ for the asymmetric lattice with open boundary and $G(r)$ for the symmetric lattice coincides with each other, while $G_{\parallel}(r)$ for the asymmetric lattice with magnetized boundary is suppressed.

\begin{figure}[H]
\centering
\includegraphics*[ width = 0.7\textwidth]{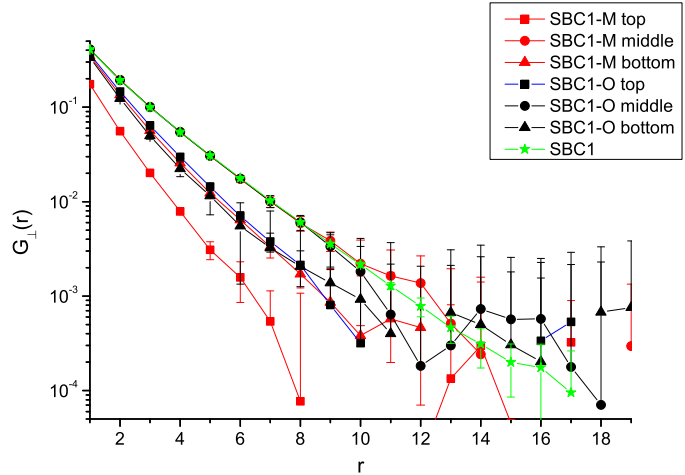}
\caption{Transverse spin-spin correlation $G_{\perp}(r)$ between spins the same distance $r$ from the boundary of a $100 \times 100$ Ising lattice at $T = 3.0$.  In this figure, $G(r)$ of the symmetric lattice with no special boundaries is shown in green, $G_{\perp}(r)$ of the asymmetric lattice with a magnetized boundary are shown in red, while $G_{\perp}(r)$ of the asymmetric lattice with an open boundary are shown in blue.}
\label{fig:olr}
\end{figure}

Finally, we show in Figure \ref{fig:olr} the transverse spin-spin correlation function $G_{\perp}(r)$ between spins at the same distance $r$ from the boundary, for a $100 \times 100$ Ising lattice at $T = 3.0$.  Here we see that $G_{\perp}(r)$ measured in the middle of the lattice converge to $G(r)$ of the symmetric lattice for both magnetized and open boundary conditions, i.e. the effects of the boundary conditions, whatever they are, are negligible this far away.  For the row of spins adjacent to the stochastic boundary opposite the special boundary, $G_{\perp}(r)$ is suppressed by the same extent for both boundary conditions.  Since these spins are all adjacent to SBC1 pseudospins, we judged that this common suppression is due to the `imperfect' SBC1 algorithm.  More importantly, for the row of spins adjacent to the special boundary, $G_{\perp}(r)$ is suppressed beyond the SBC1 imperfection only for the asymmetric lattice with a magnetized boundary.

\section{Conclusions}
\label{sect:conclusions}

To summarize, we introduced in this paper a hierarchy of stochastic boundary conditions (SBCs) for Monte Carlo simulations of the two dimensional Ising model.  The main idea behind our SBCs is to couple spins at the boundaries of the system to independent pseudospins, whose dynamics and local correlations mimick spins in the bulk of the system.  We then described the zeroth (SBC0) and first (SBC1) order algorithms in detail, in particular how we collect spin flip statistics in a PBC calibration stage, before turning on SBC in the implementation stage, and allowing the simulation to achieve self-consistency before measurements are performed.  At simulation temperatures above $T_C$, we find that self-consistency is reached very quickly.  From measurements of the specific heat, magnetic susceptibility, and spin-spin correlations, we found the two SBCs compare favorably against the PBC, and that SBC1 systematically improves upon SBC0.

We then demonstrate the advantage of using SBCs to simulate two asymmetric Ising lattices, one with a magnetized boundary, and another with an open boundary.  To do this, we allow the SBC to vary with distance from the special boundary.  We then checked from measurements of row magnetization per spin and transverse spin-spin correlations that the modified SBC0 and modified SBC1 agree with each other, and also with the bulk results expected far from the special boundary.  To understand the effects of the magnetized and open boundaries, we measure the row magnetic susceptibility per spin, the longitudinal spin-spin correlations, and the transverse spin-spin correlations.  We find that the magnetized and open boundaries both suppress the row magnetic susceptibility near them, with the magnetized boundary more strongly so.  For the longitudinal and transverse spin-spin correlations, only the suppression near the magnetized boundary can be clearly seen in the SBC simulation results.

\section*{Acknowledgements}

This research is supported by the Nanyang Technological University grants SUG 19/07 and RG 22/09.  YW acknowledges support from the Nanyang Technological University's Undergraduate Research on Campus (URECA) program.  We have had helpful discussions with Liu Zheng, Leaw Jia Ning and Huynh Hoai Hguyen.

\end{document}